\documentclass[%
 reprint,
superscriptaddress,
nofootinbib,
 amsmath,amssymb,
 aps,
 longbibliography,
floats,
floatfix,
]{revtex4-1}

\usepackage[utf8]{inputenc} % allow utf-8 input
\usepackage[T1]{fontenc}    % use 8-bit T1 fonts
\usepackage[colorlinks=true,urlcolor=blue,linkcolor=blue,citecolor=blue]{hyperref}       % hyperlinks
\usepackage{url}            % simple URL typesetting
\usepackage{booktabs}       % professional-quality tables
\usepackage{amsfonts}       % blackboard math symbols
\usepackage{nicefrac}       % compact symbols for 1/2, etc.
\usepackage{microtype}      % microtypography
\usepackage{natbib}
\usepackage{textgreek}
\usepackage{graphicx,caption}
\usepackage{wrapfig}
\usepackage[toc,page]{appendix}
\usepackage{subcaption}
\usepackage{todonotes}
\usepackage{afterpage}

\usepackage{multirow}
\usepackage{siunitx}
\usepackage[final]{changes}

\usepackage{float}
\makeatletter
\let\newfloat\newfloat@ltx
\makeatother

\usepackage{algorithm}% http://ctan.org/pkg/algorithms
\usepackage{algpseudocode}% http://ctan.org/pkg/algorithmicx

\begin{document}

\preprint{APS/123-QED}

%\title{Masked Particle Modeling: \\ Towards High Energy Physics Foundation Models with Masked Modeling on Sets}

\title{Masked Particle Modeling on Sets: \\ Towards Self-Supervised High Energy Physics Foundation Models}
\author{Tobias Golling}
\affiliation{University of Geneva}
\author{Lukas Heinrich}
\affiliation{Technical University of Munich}
\author{Michael Kagan}
\affiliation{SLAC National Accelerator Laboratory}
\author{Samuel Klein}
% \email{samuel.klein@unige.ch}
% \thanks{Corresponding author}
\affiliation{University of Geneva}
\author{Matthew Leigh}
\affiliation{University of Geneva}
\author{Margarita Osadchy}
\affiliation{University of Haifa}
\author{John Andrew Raine}
\affiliation{University of Geneva}

\begin{abstract}
   We propose \textit{masked particle modeling} (MPM) as a self-supervised method for learning generic, transferable, and reusable representations on unordered sets of inputs for use in high energy physics (HEP) scientific data. This work provides a novel scheme to perform masked modeling based pre-training to learn permutation invariant functions on sets. More generally, this work provides a step towards building large foundation models for HEP that can be generically pre-trained with self-supervised learning and later fine-tuned for a variety of down-stream tasks. In MPM, particles in a set are masked and the training objective is to recover their identity, as defined by a discretized token representation of a pre-trained vector quantized variational autoencoder. We study the efficacy of the method in samples of high energy jets at collider physics experiments, including studies on the impact of discretization, permutation invariance, and ordering. We also study the fine-tuning capability of the model, showing that it can be adapted to tasks such as supervised and weakly supervised jet classification, and that the model can transfer efficiently with small fine-tuning data sets to new classes and new data domains.
   %Some wonderful abstract here about foundation models for HEP, tokenization, self-supervised learning with Masked particle modeling. Also maybe something about data based training and domain shift mitigation. 
\end{abstract}

\maketitle

\section{Introduction}
While Artificial Intelligence (AI) and Machine Learning (ML) are already playing a major role in the analysis of high energy physics (HEP) data, the HEP community has yet to benefit from the self-supervised learning (SSL) based approaches to building large \textit{foundation models} (FM)~\cite{bommasani2022opportunities} that have been pioneered in natural language processing (NLP)~\cite{lewis2019bart,devlin2019bert,openai2023gpt4,NEURIPS2020_1457c0d6} and computer vision (CV)~\cite{dosovitskiy2021image,CaronTMJMBJ21,bao2022beit}. {FMs, as opposed to task specific ML models, are pre-trained in generic ways, such that they are useful for a range of downstream tasks}. {FMs often} use SSL to pre-train models on vast data sets in order to learn generic representations of the data.  Such models can then be efficiently fine-tuned with small datasets for a variety of downstream tasks. The self-supervised pre-training of a FM produces a model that is also referred to as the ``backbone'', as it can serve as the information extraction component for downstream models.  This concept significantly expands the possibilities for learning robust and meaningful data representations. The knowledge encoded in this way can be readily applied in downstream tasks which use this representation as input. 
%Fine-tuning the saved weights of a pre-trained backbone model can lead to better performance than training the model ``from scratch'' and often requires significantly less training data. \ro{a bit of a repetition about fine-tuning of on a smaller data.}

The FM with a SSL pre-training approach offers important advantages for HEP. Unlike supervised learning, which typically acquires limited domain representations and focuses on a few key features for high prediction accuracy that must be learned anew for each task, SSL aims to learn generic representations summarizing domain features that prove useful across various downstream tasks. SSL tasks can be formulated on unlabeled data. In the HEP context, this may not only decrease the reliance on labeled simulated data sets but also potentially helps mitigate uncertainties related to domain shift when training models on imperfect simulations.
However, this approach also has several major challenges for HEP. SSL strategies are data type specific, so new methods must be developed. These models also represent a scale in both model size and data size that have not been addressed in HEP. In this work, we aim to take the first steps towards building such a HEP foundation model, focusing on developing HEP data specific SSL strategies{, understanding how these models perform when fine-tuned for various downstream tasks, and} keeping an eye on how well such strategies may scale in the future. We propose a \textit{masked particle modeling} (MPM) scheme, akin to masked language modeling (MLM) in NLP, for self-supervised learning on unlabeled data consisting of sets of particles in a collider physics environment. In doing so, we propose a novel scheme to apply masked modeling strategies to unordered sets of inputs.

\begin{figure*}
\centering
\includegraphics[width=0.8\linewidth]{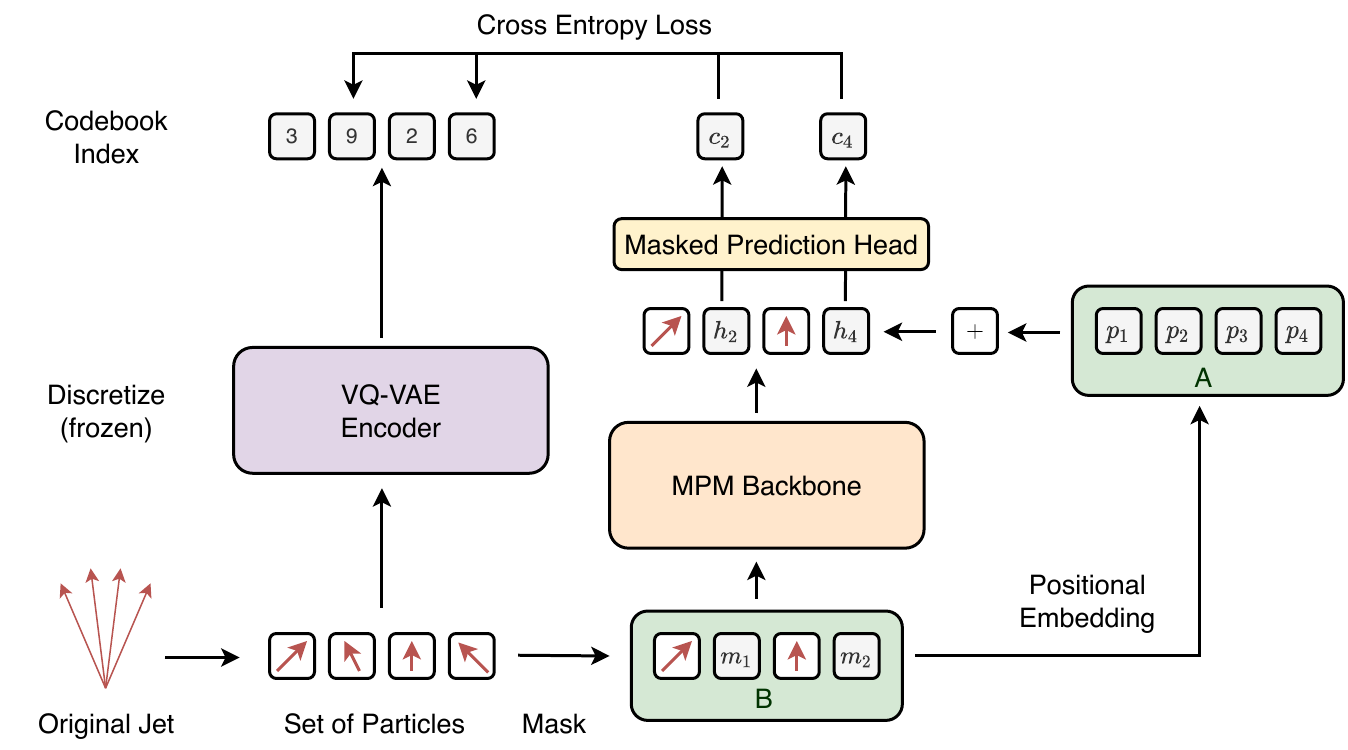}
\caption{The proposed model and training scheme for a FM for jets. A jet is represented as a set of particles, each a list of features, and some particles are replaced by a learnable vector and passed through a transformer encoder. Training aims to predict the discrete token identity, defined by the encoder of a pre-trained VQ-VAE, of the masked particles. {The MPM Backbone in is a transformer encoder in this work. The positional embedding shown in box A is used only in the prediction head to preserve the permutation invariance in the backbone\added{;} it is required in the model to break the degeneracy from when $m_1=m_2$.}
    \label{fig:mpm}   
    }
\end{figure*}

This work aims to generalize the language-inspired MLM-type training scheme to HEP scientific data, {and to show that in the HEP context this can be an effective strategy for developing a useful FM capable of being fine-tuned for various down-stream tasks.} %The paradigm is to extract semantic meaning and {\it understanding} of the whole from individual input elements together by attempting to predict what (masked) pieces, referred to as tokens, are missing. 
The paradigm involves extracting semantic meaning and understanding of the whole by predicting the missing (masked) pieces, referred to as tokens, thereby considering the collective impact of individual input elements.
Generic data sets, such as the ones in HEP, cannot be directly mapped to a sequence of pieces in analogy to words in a sentence as they are continuous and unordered, i.e. not a sequence. The challenges of using continuous elements, i.e. features of particles like momentum, instead of a discrete dictionary of words, and of operating on unordered sets of inputs will be addressed in the development of the MPM scheme. To do so, we examine MPM in collider physics data sets consisting of jets, or collimated streams of particles produced by high-energy quarks and gluons.

% {
% In this work we take the definition of a FM to be a model that is useful for a range of downstream tasks and we show that SSL can be used to construct FMs for HEP data satisfying this definition. 
% The aim of this work is not to produce the best possible SSL strategy for defining FMs, but rather to demonstrate that such models can be constructed.
% The strategy for training this model can be selected using a simple recipe and still result in a useful model.
% We also show that pre-training strategies based on masking can be transferred to HEP, such strategies have not been applied in this context before and could allow models to be trained on real data directly.
% }

We provide a brief overview of related work in Section~\ref{sec:related}. We describe the MPM method, including the token creation scheme, and the fine-tuning procedures in Section~\ref{sec:methods}. Experiments on different tokenization schemes, fine-tuning for jet classification including on unseen classes, on new datasets, and in a weakly supervised setting are presented in Section~\ref{sec:exp}.

\section{Related Work}\label{sec:related}
Foundation models, such as Masked Language Models (BART~\cite{lewis2019bart}, BERT~\cite{devlin2019bert}), Generative Pre-trained Transformer (GPT)~\cite{openai2023gpt4,NEURIPS2020_1457c0d6}, Vision Transformer (ViT)~\cite{dosovitskiy2021image}, DINO~\cite{CaronTMJMBJ21} and their combinations, such as DALLE~\cite{ramesh2021zeroshot}, Flamingo~\cite{AlayracDLMBHLMM22} and others have primarily been explored in the domains of language and vision. We refer readers to the recent review~\cite{bommasani2022opportunities} for an overview. Most closely related to this work is the BERT model~\cite{devlin2019bert}, which uses the masking and prediction of missing words as a pre-training task, and the  BEiT model~\cite{bao2022beit}, which adapts the masked language modeling method to images by masking and predicting patches of input images. On masked modeling schemes for data which consists of unordered sets of inputs, the impact of removing positional information in masked image modeling was examined in Ref.~\cite{Chen_2021_ICCV}, and using position as a target when processing unordered image patches were explored in Ref.~\cite{pmlr-v162-zhai22a}. The first steps in developing foundation models for science have been developed in e.g. protein biology~\cite{ESM}, molecular chemistry~\cite{LLM_chem1,LLM_chem2}, and cosmology~\cite{lanusse2023astroclip,walmsley2022galaxy}, showing their ability to learn informative representations that are useful in these domains for various downstream tasks.

The first steps in self-supervised learning on jets was explored in Refs.~\cite{10.21468/SciPostPhys.12.6.188,dillon2023anomalies,Tombs_2022}, largely focusing on contrastive pre-training using augmentations of jets. Pre-training strategies through masking particle type information have also been explored in Ref.~\cite{kishimoto2023pretraining}. Transformer models were trained on large jet datasets for classification in a supervised setting in Ref.~\cite{parT,Mikuni_2021} and several transformer-based applications have since been developed (for example, see Refs.~\cite{kach2022point,Kansal_2023,Fenton_2022,xbbtagger,smith2023differentiable,tomiya2023equivariant,kach2023attention,raine2023nu2flows}). Transformers have also been used for auto-regressive density estimation and jet classification~\cite{finkeQCD2023,butter2023jet}. Notably, Ref.~\cite{finkeQCD2023} also explored the discretization of continuous particle features to form jet sequences, which we examine in this work.

In parallel to the present effort on self-supervised foundation models, investigations are ongoing on the potential of \emph{supervised} FMs in HEP by using physics-motivated pretext tasks followed by fine-tuning in a hierarchical setting~\cite{supervisedfinetuning}.

\section{Overview of Methods}\label{sec:methods}

The proposed model and training scheme is summarized in Fig.~\ref{fig:mpm}. In line with the MLM framework employed by BERT~\cite{devlin2019bert}, the MPM objective described in Section~\ref{subsec:mpm} involves selecting a subset of particles within each jet to form the masked set. A predefined masking strategy is applied to this subset. The goal of MPM is to build a model capable of inferring the attributes of the original particles within the masked set, using information from all other particles present in the jet.
As particles form unordered sets, in contrast to the sequential nature of sentences, we develop a masked prediction scheme which is applicable for unordered set-based data. An additional challenge stems from the continuous nature of particle features, in contrast to the discrete dictionary found in language models but similar to the challenges of masking image patches in CV. In Section~\ref{subsec:vqvae}, we tackle this challenge employing methods akin to those used in BEiT~\cite{bao2022beit}. We discuss the fine-tuning of the pre-trained model to downstream tasks in Sec.~\ref{subsec:finetune}

\subsection{Masked Particle Modeling}\label{subsec:mpm}

%The representation of a jet as a collection of particles, each particle being described by a set of continuous features, lends itself to the interpretation within the masked training paradigm akin to MLMs, as each particle in the jet is similar in concept to a token in a block of text (with the notable difference that particles form unordered sets, while text are sequences).
The representation of a jet as an unordered set of particles, each characterized by an ordered collection of continuous features, lends itself to interpretation within the masked training paradigm reminiscent of MLMs. In this analogy, each particle within the jet can be viewed as a representation akin to a token in a block of text, despite the notable difference that particles form unordered sets, while text is sequential.

The (MPM) objective relies on the selection of a subset of particles within a jet. This involves removing the information associated with these particles and replacing it with a learnable mask. Subsequently, the goal is to predict a certain property for each of the originally masked particles.
Modeling masked particles is anticipated to serve as a valuable pre-training task. This is because the resulting model is expected to understand properties of jets by acquiring the ability to correct or infer missing information through the analysis of unmasked particles.

We can consider jets with at most $N$ constituent particles to be described as a set $X=\{x_i\}_{i=1}^N \in \mathcal{X}$ where each $x_i$ is some representation of a particle and $\mathcal{X}$ is the set of possible jets. 
A dataset is then a collection of $K$ jets $\{X^{j}\}_{j=1}^K $. 
The MPM objective can be phrased as partitioning the set of particles in each jet $j$ into a masked set $\mathcal{M}_x^{j}=\{x_i^{j}\}_{i\in\mathcal{M}^{j}}$ over indices $\mathcal{M}^{j}$ of masked elements, and an unmasked set $\mathcal{U}_x^j=\{x_i^j\}_{i\in\mathcal{U}^{j}}$ over indices of non-masked elements. %$\mathcal{U}^{j}=\bar{\mathcal{M}}^{j}$.  
A masking strategy is used for mapping $\mathcal{M}_x^j \rightarrow \mathcal{M}_m^j = \{m_i^j\}_{i\in\mathcal{M}^j}$ where each $m_i^j$ is a learnable vector. For convenience, we drop the jet index $j$ as the procedure is repeated for all jets.
The goal is to define a parametric function $f_\theta : \mathcal{X} \to \mathbb{R}^{N \times d}$, which maps a jet to a $d$-dimensional representation for each of the $N$ particles in the jet, and a loss $\mathcal{L}$ such that minimizing $\mathbb{E}_{\mathcal{X}}[\mathcal{L}(\mathcal{M}_x, f_\theta(\mathcal{M}_m, \mathcal{U}_x))]$ results in a function $f_\theta$ that is useful for downstream tasks.
As the goal of the pre-training is to recover information about each masked particle, we use a per-particle loss function of the form
\begin{equation*}
    \mathbb{E}_{\mathcal{X}}\left[\frac{1}{\left|{\mathcal{M}_m}\right|} \sum_{i\in\mathcal{M}_m} \mathcal{L}\left(x_i, f_{\theta,i}(\mathcal{M}_m, \mathcal{U}_x)\right)\right],
    \label{eq:loss_decomp_bert}
\end{equation*}
where $f_{\theta,i}$ is the model output for the $i$th particle.

Unlike language models, which operate on a finite and discrete vocabulary of words, many of the features one that describe particles are continuous, such as momentum, direction, distance of closest approach to the primary collision, etc. This distinction has implications in two aspects of the model.  First, concerning the model input, one may choose to use the continuous features directly, following the approach typically employed in BEiT-type models. Alternatively, one could opt to discretize particle features, for example by binning along each feature dimension, and use the feature bin index as an input token for the model, as demonstrated in Ref.~\cite{finkeQCD2023}. Second, at the model output, where the prediction of missing particle features occurs, one can either directly predict continuous features or, if particle features have been discretized, employing the discrete tokens as prediction targets. 
This second choice has implications for the pre-training loss function. Predicting continuous features generally involves a regression-type loss, leading to learning the mean of the feature posterior conditioned on the unmasked particle features. On the other hand, predicting discrete indices allows pre-training to be framed as a classification problem, where the model learns a categorical posterior distribution over indices. The ability to model the full posterior and any potential multi-modality in the distribution can be highly beneficial, potentially preventing the model from allocating resources to irrelevant and excessively detailed features. 
We compare both discrete and continuous inputs to the model, and regression and classification losses.

The particles that form a jet are permutation invariant, and so it is natural to use a backbone model that is permutation equivariant.
However, if the same learnable vector is used to replace all of the particles in the masked set, then the output of the model will be equivalent at every masked position.
In Fig.~\ref{fig:mpm} this would mean that if $m_1=m_2$ then $h_2=h_4$, because permutation equivariant functions can not distinguish between two input particles with the same values.
This exposes a redundancy that can only be removed by inducing an ordering, or by adding attributed connections between particles.
The redundancy makes it impossible to achieve zero error on any single prediction unless the original value of every particle in the masked set is identical.
However, the redundancy does not make MPM fruitless, as learning the density over the masked set of particles is still a useful and difficult task. 

Two strategies are examined to address the redundancy that arises from the permutation invariance of MPM:
(a) Use the transverse momentum $p_T$ to order each particle in the jet at the input to the backbone. As this breaks the permutation invariance of the backbone for all downstream tasks, it may have adverse effects on predictive performance. (b) Employ the $p_T$ to order the particles in a jet at the input to the masked prediction head in Fig.~\ref{fig:mpm}. This head is not used for downstream tasks, ensuring that this approach does not break the permutation invariance of the backbone.
Both these approaches are compared with retaining the redundancy (i.e., no ordering).

\subsection{Making Tokens From Particles}\label{subsec:vqvae}

To evaluate the impact of employing a discretized set of tokens for model input or property prediction, a suitable tokenization scheme is needed. As noted earlier, the simplest scheme involves binning each feature into a finite set of feature ranges. While binning features can be used to define a set of labels for classification targets, this method falls short in incorporating contextual information, such as information about particle features in relation to the features of other particles within the jet.

A scheme for defining context dependent tokens from continuous inputs was developed in the masked image modeling approach of the BEiT model~\cite{bao2022beit}. In this case, labels of different image patches were defined using a Vector Quantized Variational AutoEncoder (VQ-VAE)~\cite{oord2017neural}.
%Turning the continuous inputs into discrete targets allows the pretraining task to be framed as a classification problem. This was found to be important in BEiT~\cite{bao2022beit} as regressing the continuous inputs directly wasted modelling capacity on irrelevant features of the inputs.
A VQ-VAE uses an encoder to map a set of inputs to latent vectors, which are subsequently projected onto the nearest element within a finite codebook. These codebook vectors are then decoded back to the original inputs. The codebook vectors are trained simultaneously with the encoder and decoder. The use of transformer encoder and decoder ensures that information from all input elements are used to define the predictions. This process incorporates input-wide context into the codebook prediction for each element within the input set. 
% Several prescriptions have been developed to overcome these issues and we use the methods outlined in Ref.~\cite{huh2023straightening} as described in detail in App.~\ref{app:vq_vae}.
In MPM, each particle is encoded to a single codebook element, where the encoding is performed conditionally on all other particles in the jet.
The index of the codebook element to which each particle is mapped is then used as the target label during pre-training.
{The VQ-VAE model is only used during pretraining.}
% A full description of the architecture and training procedure for the VQ-VAE we used is provided in App.~\ref{app:vq_vae}.

We also explore using the K-means algorithm~\cite{macqueen1967some} to define labels for the pre-training.
We use the K-means++~\cite{kmeansplusplus} algorithm as implemented in the scikit learn library~\cite{sklearn_api} to define the clusters. After training, each cluster is assigned an index, and the index is used as target labels.
This approach is explored as a replacement for the VQ-VAE defined labels.
{Unlike the VQ-VAE labelling, the K-means approach is context independent.
Further studies into the differences between the VQ-VAE and K-means would be beneficial.
}

\subsection{Fine-Tuning}\label{subsec:finetune}

Evaluating the utility of a pre-trained backbone is a non-trivial task, since a suitable set of downstream tasks for benchmarking the performance of the pre-trained models is required. 
In this paper we focus on jet classification in different contexts.
The task of jet classification is performed by applying weighted pooling to the output of the backbone followed by a linear layer to map to the same number of dimensions as there are jet classes.  

To assess the impact of the pre-training and fine-tuning we explore three strategies to train a model for a downstream task.
The first is referred to as ``fixed backbone'', where the pre-trained backbone is frozen during fine-tuning and only a linear classification head is updated. This tests the power of the representation that is learned during fine-tuning, specifically the linear separability of different classes in this representation.
The second is referred to as ``fine-tuned'', where both a linear classification head and the backbone model itself is fit to the downstream task.
This tests the utility of the pre-training task for defining an initialization of the function parameterized by our model.
The third is referred to as ``from scratch'', where the backbone model is reinitialized with new weights and fit directly on the downstream task, i.e. standard supervised learning.
This third approach provides a performance benchmark for a given model and training strategy on a given downstream task without any influence from the pre-training.

\section{Data Sets}\label{sec:data}
We use the JetClass dataset~\cite{jetclass} for all pre-training tasks.
The dataset contains 100 million training samples from ten different classes with an equal number of samples for every class.
Each class in the JetClass dataset represents jets resulting from a specific decay chain involving different particles. 
For example, the class labelled $H \rightarrow b\overline{b}$ includes jets which result from the decay of a Higgs boson into a $b$ and anti-$b$ quark before they hadronize and decay into the many particles captured by the detector. 
Other classes include jets initiated by quarks $q$, gluons $g$, top quarks $t$, and the $W$ or $Z$ vector bosons.
The different processes are also distinguished by whether the decay chain contains a lepton $\ell$, bottom $b$ quark or charm $c$ quark. In this work, the momentum and direction (azimuth $\phi$ and pseudo-rapidity $\eta$) are used as the feature collection for each particle.

% {\color{red}The RODEM data set...}
The RODEM dataset comprises additional independent samples of top quark initiated jets and jets arising from gluons and quarks (QCD).
It should be noted that in this dataset no distinction is made between quark and gluon initiated QCD jets, unlike in the JetClass dataset.
Similar to JetClass, both samples are generated with MadGraph~\cite{MadGraph} interfaced to Pythia8~\cite{Pythia}, however in these samples the decay of top quarks and $W$ bosons is performed using MadSpin~\cite{MadSpin}.
Another difference with respect to the JetClass datasets is in the detector simulation.
Here, the Delphes~\cite{delphes} detector simulation is performed with a parameterization similar to the ATLAS detector, rather than the CMS detector parameterization used in JetClass.
Furthermore, the anti-$k_t$ jet clustering~\cite{AntiKt} is performed with a radius parameter of 1.0 rather than 0.8, and the selected jets fall in a slightly wider range of transverse momentum, spanning 450~GeV to 1.2~TeV, and an increased pseudorapidity range of $|\eta| < 2.5$.
10~million jets of each class are available, with particles represented only by their four-momenta and are assumed massless.

\section{Experiments}\label{sec:exp}
Experiments are designed both to test and define model design strategies and to explore the performance of the models on downstream tasks after fine-tuning. The ``design strategy'' experiments (or experiments studying design strategy choices) aim to understand the impact of tokenization on both the inputs and targets, to examine the quality of tokens learned by the VQ-VAE, and to test the impact of order the inputs or intermediate representations in the FM.
% To explore a wide array of fine-tuned model performance indicators, we examine how the model performs on various downstream tasks. 
We use a wide array of downstream tasks to provide multiple indicators of the utility of the pre-trained models.
This includes (a) in-context prediction: making predictions when using the same data set and the same classes seen in pre-training, (b) out-of-context prediction: making predictions when using the same data set at pre-training but on classes not seen in pre-training, and (c) out-of-domain prediction: making predictions using a different dataset to the pre-training, which potentially includes a distribution shift. We examine these fine-tuned performance metrics as a function of the size of the fine-tuning labelled dataset to understand how pre-training may reduce the quantity of labelled data needed for fine-tuning.

The same transformer encoder architecture is used for all backbone models. We used transformer blocks based on the Normformer~\cite{shleifer2021normformer}. Eight transformer blocks are used with a model dimension of $1024$, the input nodes are embedded linearly into a $512$ dimensional space, and the model contains a total of 40 million parameters. 
The pre-training head is a single transformer block followed by a softmax for classification or a linear layer for regression. The fine-tuning classification head is a weighted average over all output dimensions (i.e. a linear transform applied independently to each particle representation and then averaged) followed by another linear layer and softmax. Models are trained with \textsc{AdamW}~\cite{kingma2014adam,loshchilov2017decoupled} with a learning rate of $10^{-4}$ and weight decay of $0.01$. 
All pre-trained models are trained for five epochs on the full JetClass data set. For supervised training, all models are trained for up to $50$ epochs with early stopping and a patience of five epochs.
The norm of the gradient vectors are clipped to five.
The from scratch supervised models use the same architecture as the pre-trained models: eight transformer blocks followed by a weighted average and softmax for classification. 
The model architecture is similar to that used by ParT~\cite{parT}, which is state of the art. The ParT model uses more features than we use in this work, and so achieves higher accuracy.
More details can be found in App.~\ref{app:mpmarch}.

All experiments are repeated five times with different random seeds.
The uncertainty in results is given by the standard deviation across all runs, with the mean indicating the average behaviour.
The uncertainty that comes from pre-training models is not included, i.e. models are pre-trained only once.

To define target labels and for input quantization, 512 possible token values are used. This corresponds to 512 vectors in the VQ-VAE codebook, or to 512 clusters for the K-means.

All the code used to run these experiments is publicly available at \href{https://github.com/rodem-hep/mpm}{github.com/rodem-hep/mpm}.
% {All the code used to run these experiments is publicly available at \href{https://github.com}{github.com}.} 

\subsection{Token Creation with a VQ-VAE}
One challenge with training VQ-VAEs is the possibility of codebook collapse, where only a few of the codebook vectors are used to encode particle features. In the case of collapse to a single vector, the model would effectively be learning an average representation, which would not aid pre-training. 
To address this the prescriptions outlined in Ref.~\cite{huh2023straightening} are applied as described in App.~\ref{app:vq_vae}.
After training we find that the codebook has $\sim80\%$ utilization.
{The quantized vectors, defined by their index or codebook vector, will be referred to as tokens.}

\begin{figure}[htp!]
\centering
\includegraphics[width=0.98\columnwidth]{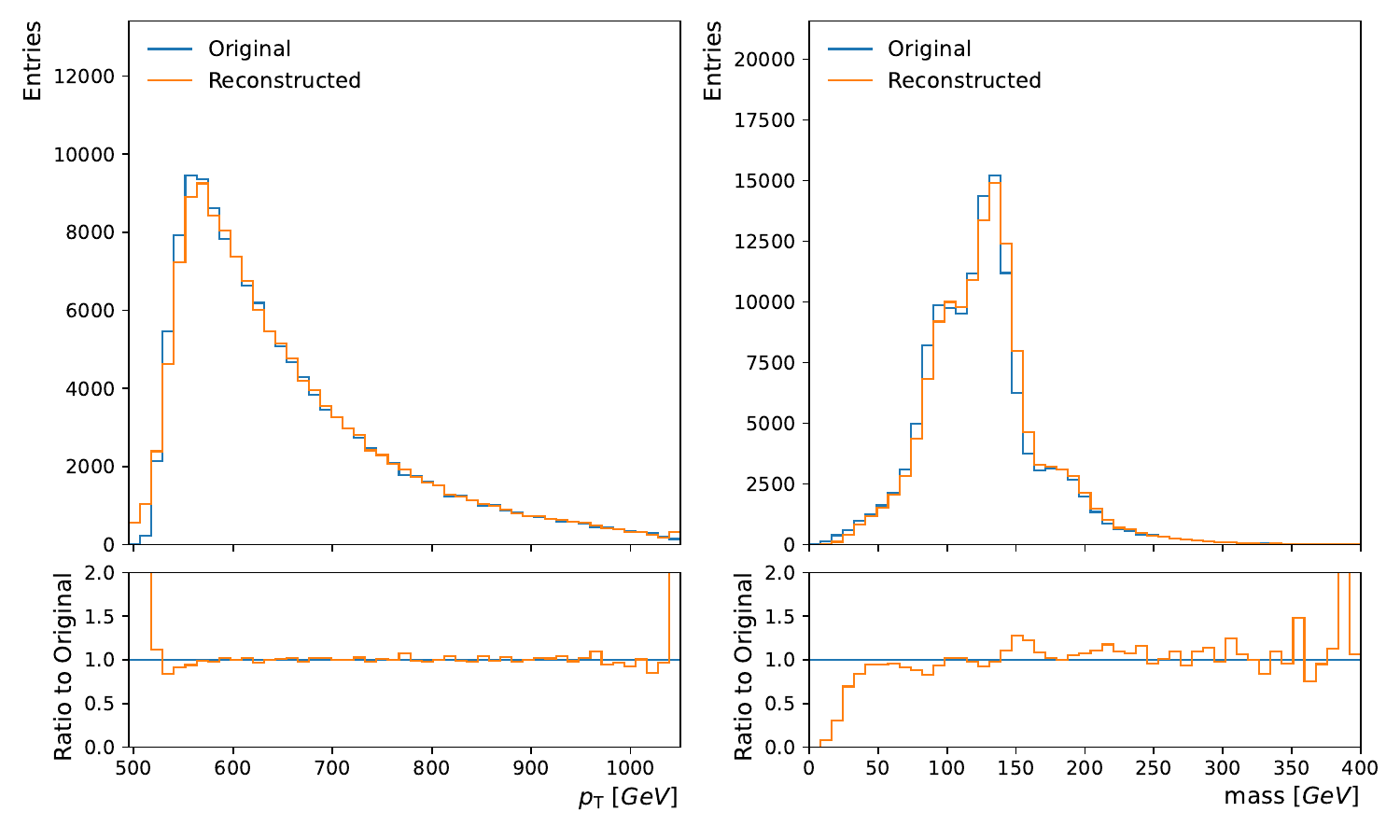}
\caption{
    \label{fig:vq_vae} 
    The reconstruction of the jet mass and transverse momentum using the decoded output of the VQ-VAE trained to quantize the particles in a jet.
    }
\end{figure}

To ensure that the latent space codes are sufficiently capturing the salient information about the particles in a jet, we examine the quality of the decoded jet properties after encoding and quantization in the VQ-VAE. Successful reconstruction of the input jet would indicate a strong performance of the encoding. In Fig.~\ref{fig:vq_vae} we show the distributions of decoded jet transverse momentum and mass are in good agreement with the input distributions. One may note the apparent overlap in goals between the VQ-VAE and the MPM model, however VQ-VAEs can be seen as primarily learning a data compression~\cite{oord2017neural} while masked particle models learn to understand a jet by inferring missing information.

\subsection{Discretization and Ordering}
In this section we explore different choices that need to be made when pre-training the model. We study the impact of ordering, input quantization, and output quantization / loss function choice. Backbone models are pre-trained under these different settings.
The possible ordering settings included are: not ordering anywhere in the model, ordering the input to the backbone, and ordering only at the input to the pre-training prediction head. 
Ordering is defined by decreasing particle transverse momentum, and the index within the ordered sequence is embedded using learned positional embeddings~\cite{gehring2017convolutional,arxiv-1706.03762}. Learned positional embedding assigns a learnable vector to each index value that is then added to the features (and thus has the same dimension as the features per element)\added{ and updated with the rest of the models parameters.}
% {The index value is set by the $p_T$ order of the constituent in the jet.}
Depending on the model  under study, a learned positional embedding is added to the backbone input features in box B of Fig.~\ref{fig:mpm} or the prediction head input features as shown in box A of Fig.~\ref{fig:mpm}.
Continuous inputs are used in all models except one, which tests the impact of quantizing particle feature inputs. 
For this input-quantized case, the codebook vectors assigned by the VQ-VAE are used to replace the input vectors from the data set in box B of Fig.~\ref{fig:mpm}. 
Finally, the classification loss using VQ-VAE encoded tokens as targets, the classification loss using K-means encoded tokens as targets, and the direct particle feature regression are tested. 

The different settings are compared by using the classification accuracy over the ten classes in a hold-out validation JetClass dataset of one million jets. Classification is performed with a linear classification head on top of the backbone output representations. Here the backbone is fixed and accuracy is compared after 20 epochs of training the linear classification head, where it is observed that the loss functions only change slowly upon further training.  

The results of the comparison can be seen in Tab.~\ref{tab:pretraining_compare} where five different random seeds were used during the fine tuning stage, but with only a single backbone trained in each case. The mean result is shown in the table, with a standard deviation of O(0.01\%) for each result. A clear benefit to ordering only in the model head is observed over ordering the backbone inputs or not ordering anywhere in the model. In this case, ordering the prediction head enables the symmetry over masked elements to be broken and thus make more precise per-particle predictions rather than a joint posterior over all masked particles. 
Input quantization is observed to hurt model performance, owing to the loss of information when summarizing continuous features with discrete tokens.  Using tokens, either from the VQ-VAE or the K-means, as classification targets is found to be substantially better than regressing particle features as a prediction task. This is likely due to the regression giving only a point prediction of average particle properties under the posterior, which provides little predictive power for posteriors with multiple modes. The classification loss is much more flexible and able to make use of the prediction of the full posterior over tokens. The VQ-VAE tokens outperformed the K-means tokens for classification, likely because the VQ-VAE is a more complex model able to capture more subtle details. Nonetheless, the K-means tokens for classification work reasonably well, and may be a useful avenue to explore further in future work due to the relative ease of training the K-means in comparison to the VQ-VAE.

Following these results, all subsequent tests use a backbone with continuous inputs, ordering in the prediction head for pre-training and a classification loss using VQ-VAE tokens as targets in the pre-training.
Unless explicitly stated the same backbone is used for all tests, this backbone is trained on the full JetClass training set using the VQ-VAE provided labels.
\replaced{We note that using a different head than linear could result in a different model selection}{We note that using a different head than linear could result in a different model selection.} However using a linear head provides the benefit of focusing our examination only on the learning in the backbone, as is often done in backbone training exploration, and enables us to select model backbones that perform better than training from scratch.

\begin{table}[htp!]
    \centering
    \caption{Linear probe into the performance of different pre-training strategies.
    For each pre-training strategy the accuracy on the ten classes of JetClass is reported.
    The accuracy is calculated over five different runs of the fine tuning with errors on the order of $0.01\%$
    }
    \resizebox{\columnwidth}{!}{
    \begin{tabular}{|l|l|l||c|}
         \hline
         Ordering & Inputs & Loss & Accuracy \\
         \hline \hline
         no ordering & continuous & VQ-VAE classification & $54.1\%$ \\
         \hline
         \textbf{order head} & \textbf{continuous} & \textbf{VQ-VAE classification} & $\mathbf{56.8\%}$ \\
         order backbone & continuous & VQ-VAE classification & $53.4\%$ \\
         \hline
         order head & quantized & VQ-VAE classification & $51.1\%$ \\
         \hline
         order head & quantized & K-means classification & $49.3\%$ \\
         order head & continuous & K-means classification & $56.2\%$ \\
         \hline
         order head & continuous & regression & $48.9\%$ \\
         order backbone & continuous & regression & $46.3\%$ \\           
        \hline
    \end{tabular}
    }
    \label{tab:pretraining_compare}
\end{table}

\subsection{Fine-Tuning for Jet Classification}
\begin{figure}[tp!]
\centering
\includegraphics[width=0.9\columnwidth]{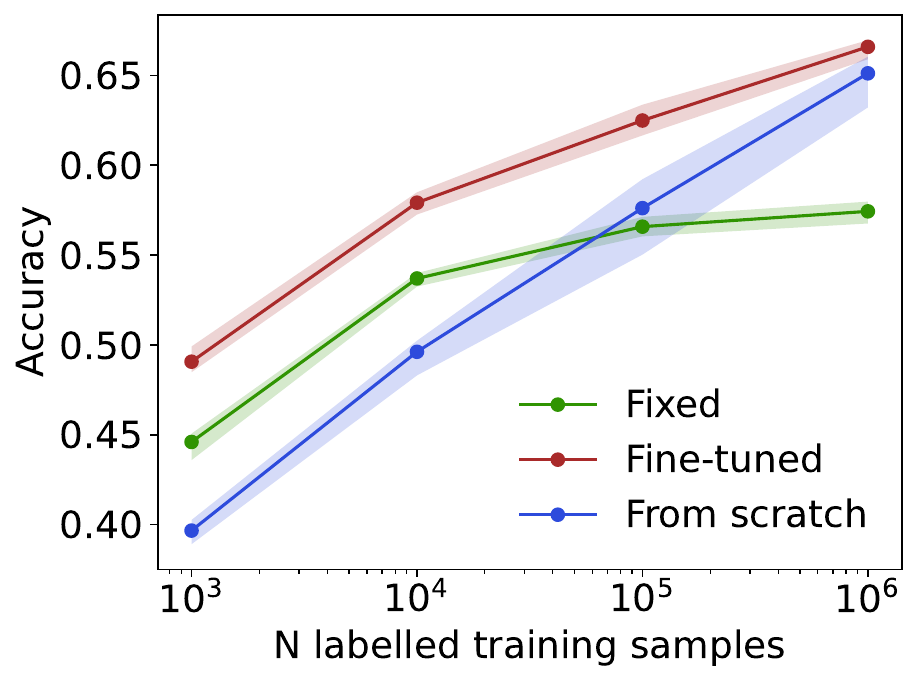}
\caption{
    \label{fig:fine_tune_jetclass} Accuracy of different training strategies as a function of the number of labelled training samples. Accuracy is calculated on the ten classes in the JetClass dataset. The average and standard deviation over 5 trainings is shown in solid lines and uncertainty bands, respectively.
    {Models with frozen pretrained backbone weights during fine-tuning are ``Fixed'', and those with updated weights are ``Fine-tuned''.}
    }
\end{figure}

The first test of the utility of the backbone model for downstream tasks is on in-context data, where we examine the accuracy of ten-class classification on a test sample from JetClass. The labelled dataset size, used for fine-tuning the pre-trained models and for training the fully supervised model, is varied to examine how much performance is gained by pre-training. 
In Fig.~\ref{fig:fine_tune_jetclass}, we can see that at small labelled data set sizes of fewer than 10k jets, there is a large performance benefit to using pre-training over the from scratch supervised model. This indicates the utility of the representation learned during pre-training for downstream tasks. With a large enough labelled data set size, the supervised model outperforms the fixed backbone model and converges to the performance of the fine-tuned backbone model. This is expected, as supervised learning on a sufficiently large labelled data set should provide enough information for model optimization without pre-training. 
In essence, the pre-training provides an excellent set of initial weights for fine-tuning on downstream tasks.

\subsection{Fine-Tuning for Jet Classification on New Classes}
\begin{figure}[tp!]
\centering
\includegraphics[width=0.9\columnwidth]{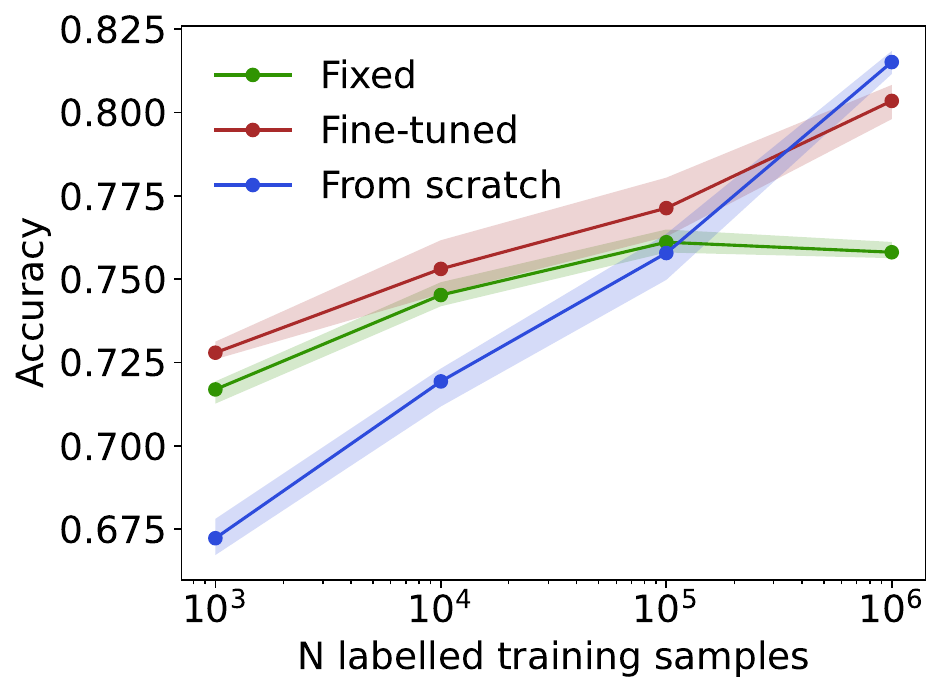}
\caption{
    \label{fig:fine_tune_jetclass_fourty} Accuracy of different training strategies as a function of labelled training samples. Accuracy is calculated on four classes, held out from all pre-trained models, in the JetClass dataset. The average and standard deviation over 5 trainings is shown in solid lines and uncertainty bands, respectively. 
    {Models with frozen pretrained backbone weights during fine-tuning are ``Fixed'', and those with updated weights are ``Fine-tuned''.}
    }
\end{figure}

To test whether the pre-trained model learns features that are generically useful for downstream tasks, or only those which are useful for the classes that are seen in the pre-training set, we perform an out-of-context test where we pre-train on a subset of the classes provided in JetClass and test the fine-tuning on the remaining classes.
Specifically, the pre-training is performed on the six Higgs and QCD classes, and then fine-tuned on the remaining four classes.
{The four remaining classes are top quarks decaying with and without a lepton, and $W, Z$ decays each as a distinct class.}
The results shown in Fig.~\ref{fig:fine_tune_jetclass_fourty}  indicate that both pre-trained models, with fixed and with fine-tuned backbone, outperform the fully supervised training when only a limited amount of data is available. This indicates that even with small amounts of labelled data, the representations learned during pre-training are generically useful for out-of-context classes. As also expected, with enough labelled training data, the fully supervised model can surpass the fixed backbone model and converge to a similar performance to the fine-tuned model. In essence, the additional labelled data allows both the fine-tuned backbone model and the supervised model to adapt the representations to the specific downstream tasks.

\subsection{Fine-Tuning for Jet Classification on New Data Sets}
We also test how well the backbone representation works out-of-domain on a different dataset of jets and how well the model can adapt to this new domain with fine-tuning. In this case, pre-training is performed with one dataset, JetClass, and fine-tuning is performed with a different data set, RODEM. More abstractly, this is a test of how well such a pre-training and fine-tuning strategy with mixed data sets may help address the domain shift challenge in HEP. This challenge come from the small, but potentially significant, differences between simulated data and real experimental data, such that models trained on simulated data have differences in prediction performance between simulated and real data, thus causing systematic uncertainties. {As such, this test is in analogy with the idea that one may want to pre-train on real data, and only fine-tune on small simulated data sets, and we hope to explore the applicability of such an approach.} With the MPM scheme, we can pre-train directly on unlabelled data, i.e. we can pre-train directly on real experimental data for representation learning. Thus we can also explore how well the features learned from data may maintain predictive power even after fine-tuning on simulation, represented here by the second labelled data set used for fine-tuning. In general, training primarily on real data has many potential benefits. It promises a reduced sensitivity to simulation-related domain shifts and calibration effects, leading to an overall simplified calibration procedure. An overall reduced simulation budget may also help alleviate the growing compute and storage limitations of the LHC experiments.
%{The test that we run here is only an analogy for this setting.}

\begin{figure}[htp!]
\centering 
\includegraphics[width=0.98\columnwidth]{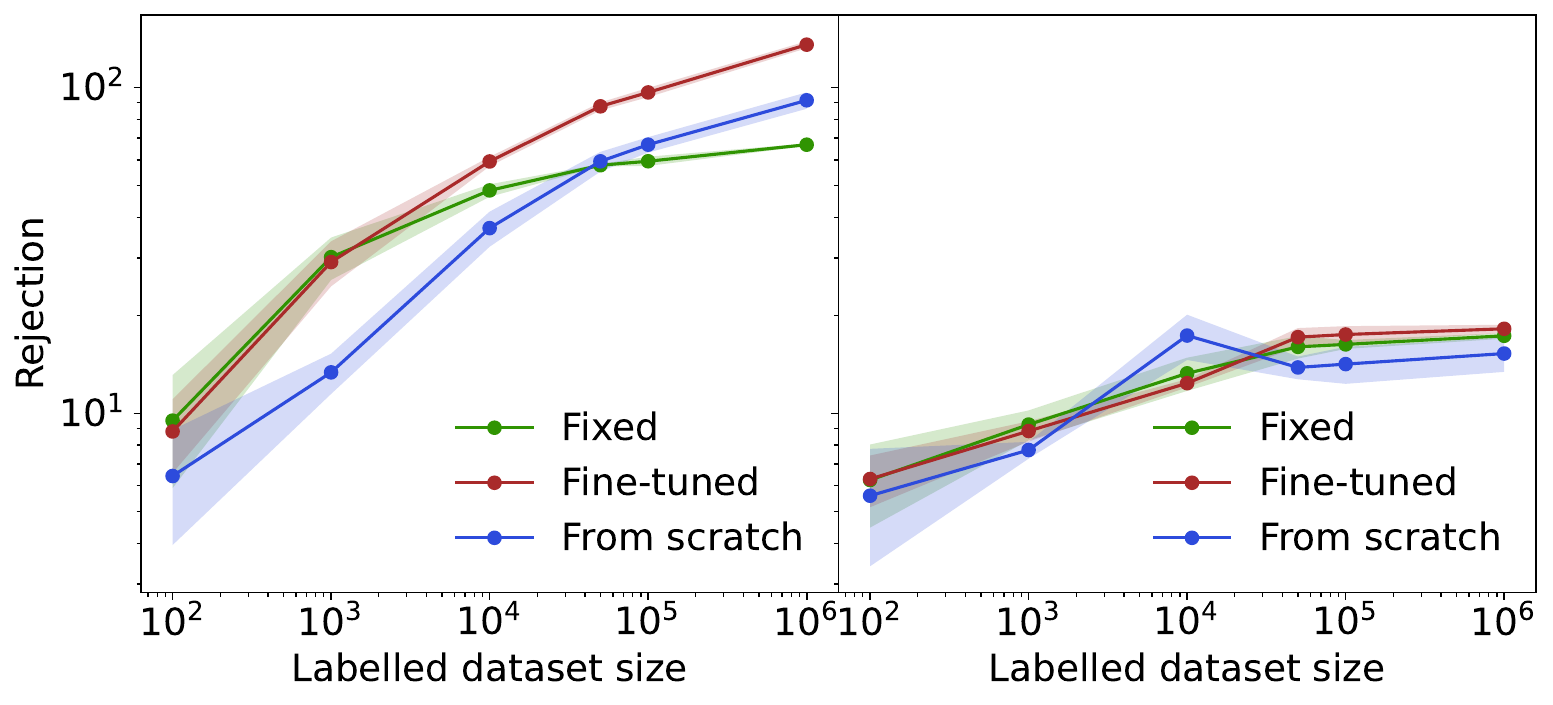}
\caption{
    \label{fig:domain_shift_ttbar} The QCD rejection at $50\%$ top-jet efficiency evaluated on (left) the RODEM test set and (right) the JetClass test set, as a function of the size of the RODEM data set used for fine-tuning. 
    All models are pre-trained on JetClass. The average and standard deviation of rejection over 5 trainings is shown in solid lines and uncertainty bands, respectively.
    {Models with frozen pretrained backbone weights during fine-tuning are ``Fixed'', and those with updated weights are ``Fine-tuned''.}
    }
\end{figure}

To perform these tests, we examine the QCD background rejection (one divided by the false positive rate) at $50\%$ top-jet signal efficiency (i.e. true positive rate) as a function of the labelled data set size. 
%We also note that a slightly smaller model of 1M parameters is used in this test. 
Fig.~\ref{fig:domain_shift_ttbar} shows the rejection for the fine-tuning RODEM data set on the left, and the rejection performance in the JetClass data set on the right. Note that the $x$-axis in both cases is the size of the RODEM labelled data set used for fine-tuning.

The pre-trained models demonstrate excellent data efficiency and performance when transferred to the RODEM dataset. The fine-tuned model outperforms the supervised model for all labelled data set sizes, as does the fixed backbone up to labelled data sets of O(50k) events.
These results further support the hypothesis that a pre-trained model learns generic and useful data representations that are not limited to the specific data set. 
This result further suggests that models can be pre-trained on real data and fine-tuned in simulation while still maintaining reasonable performance on the original real data. 
%The improvements of fine-tuning over training ``from scratch'' is more visible at small dataset sizes and bodes well for fine-tuning tasks with limited statistics.
Intriguingly, neither the fine-tuned nor from-scratch supervised model are able to outperform the fixed backbone model performance on the JetClass, i.e. ``real", data set after fine-tuning on RODEM data set.  
%Intriguingly, the model with the fixed backbone tends to do as well or better than the fully fine-tuned model on the JetClass, i.e. real data set, and exceeds performance at large labelled data set sizes. 
This suggest that the fixed features, pre-trained on the JetClass, are the most powerful for downstream tasks on the JetClass data set, while the additional representational information learned on the labelled data set provides some, but limited additional performance when applying the model on the real data. This offers a potential avenue for mitigating domain shifts in the labelled data by improving the representation learning at the pre-training stage with larger pre-training data set sizes. We leave such tests for future work.

%However, it is also evident that the models that are fine-tuned on the RODEM dataset lose performance when frozen and applied back to the JetClass dataset.
%While this is not worse than training from scratch, it does suggest that new techniques are needed to reduce this shift.

\subsection{Weakly supervised classification}
% Weakly supervised task with label proportions.
Fine-tuning tasks are typically of a supervised nature and require labels.  However, physics knowledge can be exploited to enrich data in certain classes and provide training data for fine-tuning tasks with so-called noisy labels~\cite{cwola}.  These could replace or complement fine-tuning with simulated data by leveraging pure labels in simulation with noisy labels in data.

\begin{figure}[htp!]
\centering
\includegraphics[width=0.9\columnwidth]{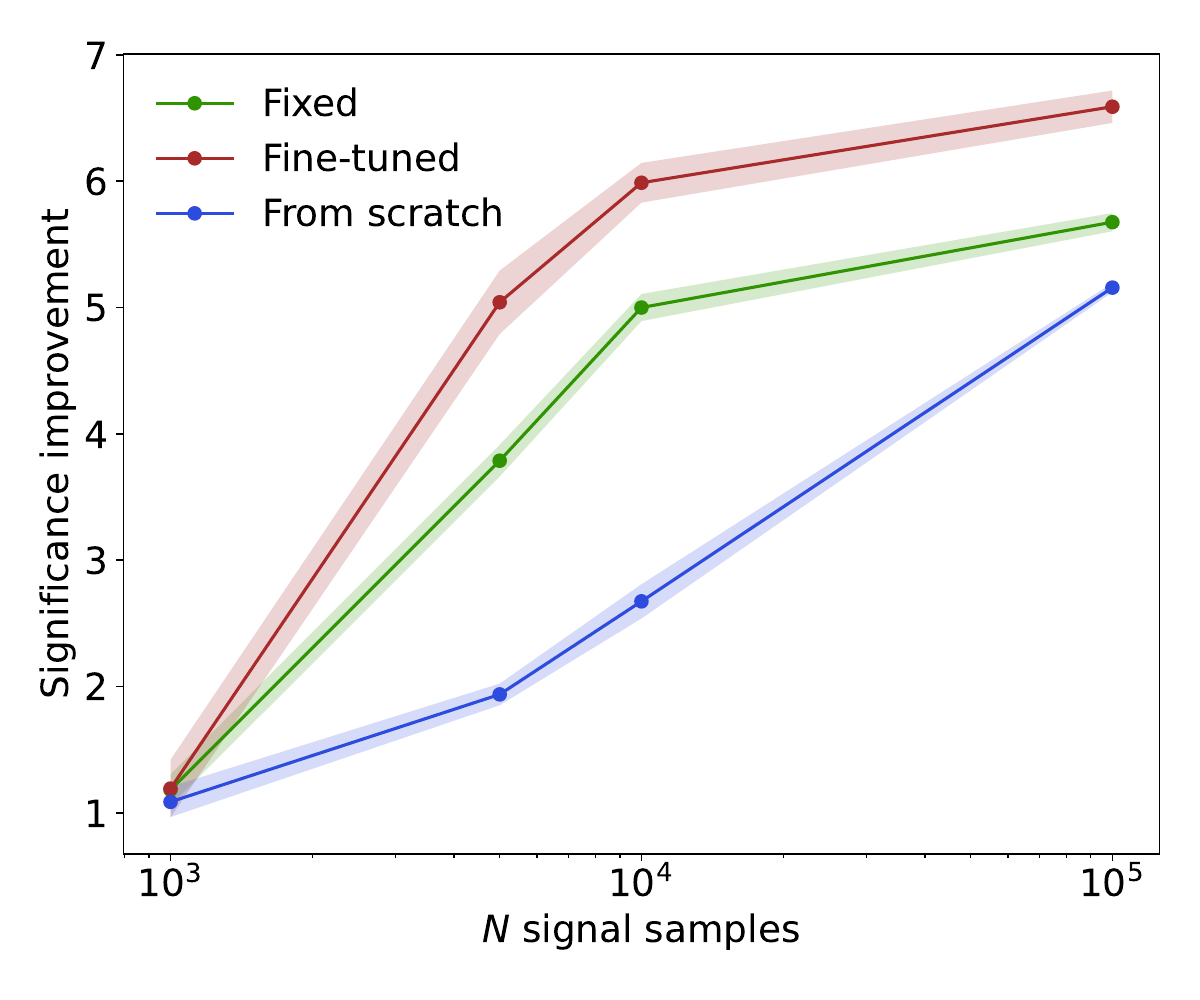}
\caption{
\label{fig:lp_ws} Models trained with weak supervision to classify data sets of different label proportions.
A data set of one million QCD jets is compared to a data set with one million QCD jets plus $N$ top jets.  The average and standard deviation of the significant improvement over 5 trainings is shown in solid lines and uncertainty bands, respectively.
{Models with frozen pretrained backbone weights during fine-tuning are ``Fixed'', and those with updated weights are ``Fine-tuned''.}
}
\end{figure}

%In this task we explore the setting where pure labels do not exist for the data, and instead we have a signal enriched data sample and a signal depleted data sample~\cite{cwola}.
This idea can be emulated by taking two samples of one class (QCD jets), each with one million events, and adding $N$ samples of another class, considered signal (top-quark initiated jets, or top jets), to one of the datasets.
The task is to then train a supervised classifier to discriminate between these two datasets and evaluate the resulting classifiers performance on separating pure datasets of each class (QCD vs top jets).  In Fig.~\ref{fig:lp_ws}, we show the significance improvement, defined as the ratio of significance before and after applying a threshold of 0.5 on the classifier output. The significance is defined as the number of signal class events divided by the square root of the number of background class events that pass a given threshold. 
Note that ground truth labels are used in the evaluation metric, but not in the fine-tuning procedure. Any value of this metric below 2 is not considered to be particularly useful. We can see that the pre-trained backbone is highly useful for this task, significantly improving the performance of the model that is trained from scratch, even when only the linear head is fine-tuned.
%As a performance metric we look at the maximum significance improvement, defined as the signal efficiency over the square root of the background efficiency.

The idea of noisy labels is useful in practice for data driven weakly supervised search strategies~\cite{10.3389/fdata.2023.899345,cathode,aad2020dijet,Andreassen:2020nkr,feta,Collins:2019jip,aad2020dijet,birman2022data}.
In particular, it has recently been demonstrated that these data driven techniques can be extended to constituent level representations of the jet~\cite{buhmann2023phase,sengupta2023improving} where the pre-training we propose here will be of significant benefit.
It has also been shown to be successful for isolating muons using data directly~\cite{witkowski2023learning}.

\section{Conclusions}

In this paper we propose the masked particle modelling strategy for pre-training models on unordered sets of inputs, and demonstrate that it is useful in the context of high energy physics. Both the continuous nature of particle features, as opposed to the discrete vocabulary typical of natural language, and the unordered nature of the data, as opposed to the sequential nature of text, are addressed to adapt masking strategies from natural language and computer vision to unordered sets of inputs, as is found in high energy physics data. When pre-training with the masked particle modeling strategy, we show that fine-tuned models can achieve high performance on downstream tasks, even using small fine-tuning data sets. These pre-trained models can be fine-tuned to discriminate classes which have not been seen during pre-training, can be adapt to new data sets, and show strong performance in weakly supervised settings. We explore the intriguing possibility to pre-train such models directly on experimental data, whilst only using simulations for fine-tuning. Such an approach may help mitigate uncertainties owing to the small distribution shifts between simulated and real data. Initial studies are promising, indicating that further examination of increasing the scale and size of pre-training data sets and backbone model may help overcome such domain adaptation challenges. More generally, this work suggests that continued exploration of self-supervised learning strategies for high energy physics data, coupled with increased data set and model sizes is a promising direction for the future development of machine learning in high energy physics.

\section*{Acknowledgements}
%\vspace{0.1cm}
MK is supported by the US Department of Energy (DOE) under grant DE-AC02-76SF00515. LH is supported by the Excellence Cluster ORIGINS, which is funded by the Deutsche Forschungsgemeinschaft (DFG, German Research Foundation) under Germany’s Excellence Strategy - EXC-2094-390783311. MO is supported by USA-Israel BSF - 2022641.
TG, SK, ML, and JR, would like to acknowledge funding through the SNSF Sinergia grant CRSII$5\_193716$ called ``Robust Deep Density Models for High-Energy Particle Physics and Solar Flare Analysis (RODEM)'', and the SNSF project grant 200020\_212127 called ``At the two upgrade frontiers: machine learning and the ITk Pixel detector''.
ML also acknowledges the funding acquired through the Swiss Government Excellence Scholarships for Foreign Scholars.

%\clearpage
%\section*{References}
%\bibliographystyle{iopart-num}
\bibliography{bibliography}
\clearpage
\appendix

\section{MPM Encoder Architecture}\label{app:mpmarch}
\begin{figure}[htp!]
\centering
\includegraphics[width=0.9\columnwidth]{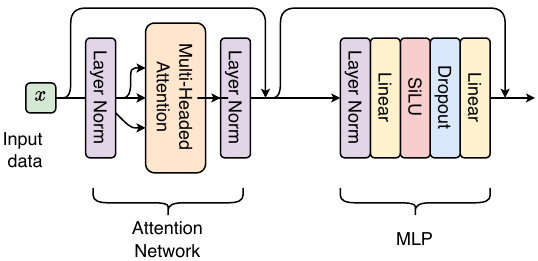}
\caption{
    \label{fig:transformer_setup} 
    The Transformer-Encoder block is made of a residual self-attention network followed by a residual dense network.
before they are passed to the dense network..  
    }
\end{figure}

The Transformer-Encoder block used in all networks is based on the Normformer~\cite{shleifer2021normformer} encoder block. It is depicted in Fig.~\ref{fig:transformer_setup}. 
The block is composed of a residual attention network followed by a residual dense network. 
The attention network takes the point cloud as input tokens and performs a multi-headed self-attention pass surrounded by layer normalizations. The intermediate tokens are then added to the input tokens via a residual connection. The dense network comprises two fully connected linear layers. A sigmoid-linear-unit (SiLU) activation is applied to the output of the hidden layer, layer normalization is used to keep the gradients stable, and dropout of $10\%$ is used for regularization. 
All models used in residual connections have zeros initialized weights in the final layer such that the total Transfomer-Encoder block is initialized to the identity.
A total of eight heads are used in the Multi-Headed Attention.
The output tokens are then added to the intermediate tokens via another residual connection. The input and output dimensions of the token features are the same, so several entire TE-Blocks can be chained together.

\section{VQ-VAE Architecture and Training}
\label{app:vq_vae}
% First define a VQ-VAE and how it is trained.
Training a vector quantized variational autoencoder (VQ-VAE) is known to be difficult~\cite{huh2023straightening}, and a useful set of prescriptions for overcoming these challenges have been outlined in Ref.~\cite{huh2023straightening}.
In the following we provide a brief overview of VQ-VAEs as defined in ~\cite{huh2023straightening}.
A VQ-VAE~\cite{oord2017neural} is composed of an encoder $F$ and decoder $G$ neural network with a quantization layer $h$ and codebook ${C}=\{c_i\}_{i=1}^m$ for vectors $c_i \in \mathbb{R}^n$ where $n$ is the dimension of the latent space.
The layer $h: \mathbb{R}^n \times {C} \rightarrow c \in {C}$ quantizes the latent space by assigning encoded vectors to their closest neighbors in the set $C$ with a distance measure defined by some measure $d$ which we will take to be the euclidean norm.
The codebook $C$ will always be omitted from the arguments of the function $h$ in the following.
The output of a VQ-VAE is defined as,
\begin{align*}
    \hat{x} &= G(h(F(x))) \\
     &= G(h(z_e)) \\
      &= G(z_q),
\end{align*}
for a given input $x$.
The objective of a VQ-VAE is to minimize the empirical risk,
\begin{equation}
    \min_{F,G,h} \mathbb{E}_x \left[ \mathcal{L}_{task}(\hat{x}, x) \right],
\end{equation}
which is not differentiable due to the quantization operation in $h$ and the gradients are estimated using straight through estimation ~\cite{bengio2013estimating}.
To ensure the accuracy of this estimation a commitment loss is added to create an attractive force between the encodings $(z_e=F(x))$ and their corresponding codebook vectors $(z_q = h(z_e))$,
\begin{equation}
    \mathcal{L}_{cmt} = (1-\beta)d(z_e,\mathrm{sg}(z_q) + \beta d(\mathrm{sg}(z_e),z_q),
\end{equation}
where $\mathrm{sg}$ is the stop gradient operator. 
The resulting fully differentiable proxy objective is
\begin{equation}
    \min_{F,G,h} \mathbb{E}_x \left[ \mathcal{L}_{task}(\hat{x}, x) + \alpha \mathcal{L}_{cmt}(x) \right],
\end{equation}
where we take take $\alpha=10$ and $\beta=0.9$~\cite{oord2017neural, huh2023straightening}.
The task loss is also taken to be the euclidean distance.

% Define the difficulties and how they were overcome

The training of this model is difficult due to well known issues with the collapse of the codebook in the latent space~\cite{huh2023straightening}. As such,
we make use of the repository from Ref.~\cite{huh2023straightening, huh2023vqtorch}.
In particular we use a shared parameterization for the codebook elements, update the commitment loss only every four steps and a synchronized update rule with $\nu=2$.
Without these additional hyperparameters we found the VQ-VAE to be very difficult to train, and we found that the VQ-VAE was not very sensitive to these parameters.

We use a model with a latent dimension of $n=16$ and $m=512$ codebook elements. 
The codebook elements are initialized using the K-means clustering algorithm on the latent space of a randomly initialized model.
The encoder and decoder networks are both transformers of the same type as those described in App.~\ref{app:mpmarch} but with four layers and a model dimension of $256$ and a linear embedding of the nodes into a $256$ dimensional space.

The input nodes are each quantized and decoded separately, with the distance calculated on a per node basis. 
Therefore a jet with $N$ constituent particles is assigned to $N$ codebook elements, each of dimension $m$.
Each of these codebook elements are decoded to $N$ vectors of the same dimension as the input nodes.
Using a transformer for the encoder and decoder networks ensures that the model is encoded and decoded conditional on all particles in the jet.
% The reconstructions of the constituents are shown in Fig.~\ref{fig:vq_vae_app}

% \begin{figure}
% \centering
% \includegraphics[width=0.9\columnwidth]{figures/csts_960.png}
% \caption{
%     \label{fig:vq_vae_app} \sk{Update this figure}.  
%     }
% \end{figure}

\section{t-SNE embeddings}

\begin{figure}
\centering
\includegraphics[width=0.9\columnwidth]{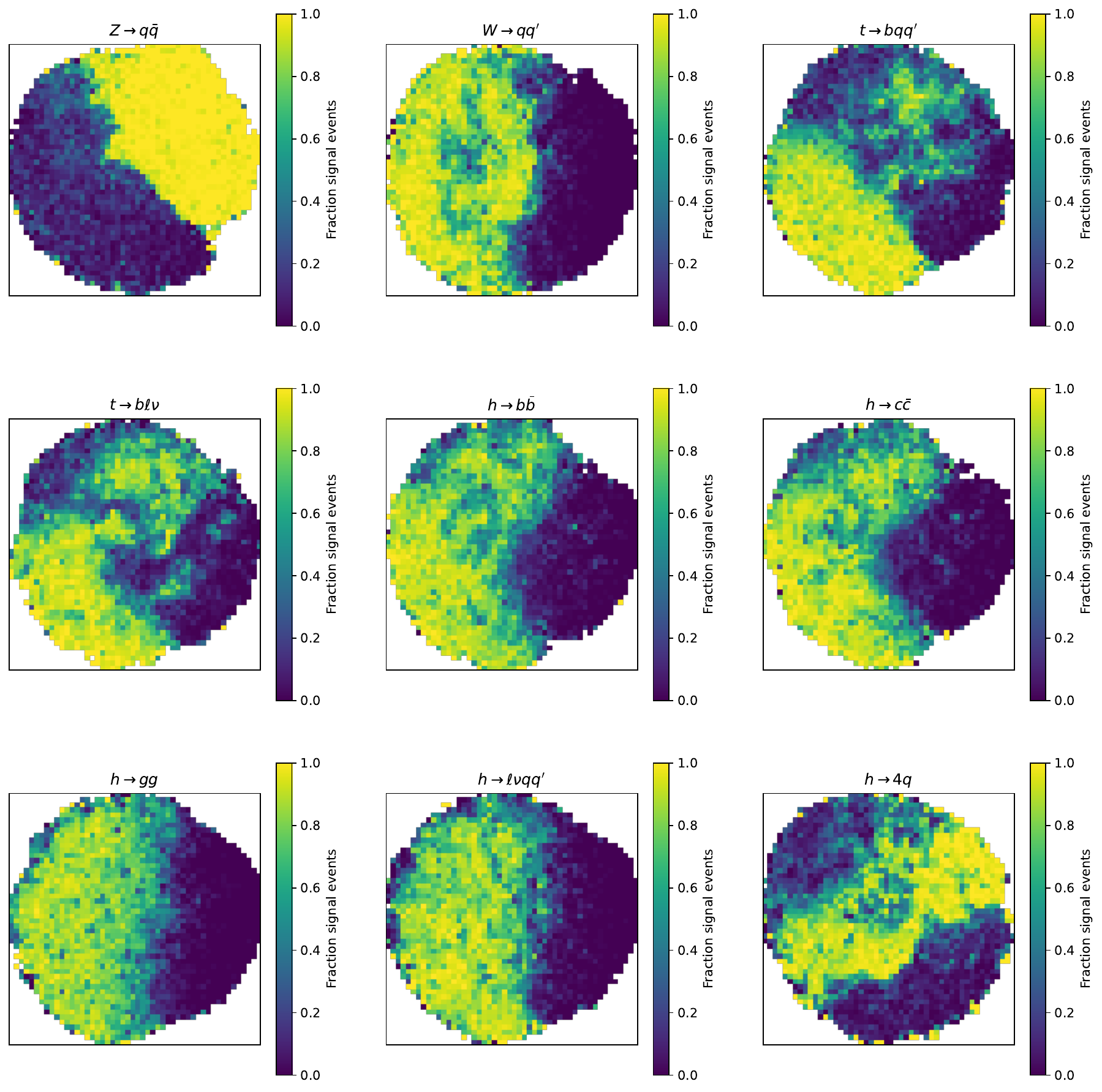}
\caption{
\label{fig:tsne_backbone} 
A t-SNE embedding of the nine signals in the JetClass dataset and the background QCD samples.
For a given signal an embedding is found for that signal and background only.
Each bin shows the fraction of signal in that bin, defined as the number of signal samples divided by the total number of samples in that bin.
For each background and signal pair, $100,000$ samples are embedded.
}
\end{figure}

In Fig.~\ref{fig:tsne_backbone} we plot the t-SNE~\cite{JMLR:v9:vandermaaten08a} embeddings at the output of the pre-trained backbone.
This shows the embedding after the self supervised pre-training.
The embeddings are performed over the representation averaged over all particles in a given jet.
Each input particle has a corresponding output representation and we embed the average over these output representations.

\end{document}